\documentclass[fleqn]{article}
\usepackage[totalwidth=7.0in,totalheight=9.25in]{geometry}

\usepackage{amsmath}
\usepackage{amssymb}
\usepackage{bm}
\usepackage{chicago}
\usepackage{etoolbox}
\usepackage[mathscr]{euscript}
\usepackage{ifthen}
\usepackage[pdftex]{hyperref}
\usepackage{manfnt}
\usepackage{mathrsfs}
\usepackage{mathtools}
\usepackage{paralist}
\usepackage{titlesec}
\usepackage{xcolor}

\newtoggle{vector-field-to-arxiv}
\newtoggle{node}
\toggletrue{vector-field-to-arxiv}
\togglefalse{node}

\newcommand{\notebeforesep}{\vspace{5pt}}

\newcommand{\nodetableofcontents}{{\begin{center}{\parbox{.95\textwidth}{\footnotesize\begin{multicols}{2}\tableofcontents\end{multicols}}}\end{center}}\vspace*{-20pt}}

\renewcommand{\cite}[1]{\citeNP{#1}}

\newcommand{\usenode}[2]{\zexternaldocument*[#1:]{../node-#1/node}\expandafter\def\csname the-#1\endcsname{#2}}

\newcommand{\noderef}[2]{\href{../#1/node.pdf}%
  {\color{refcolor}\zref{#1:#2}}~of~{\it\csname the-#1\endcsname}}

\newcommand{\nodedata}[3]{%
\expandafter\def\csname #1-title\endcsname{#2}%
\expandafter\def\csname #1-stamps\endcsname{#3}
\usenode{#1}{#2}}

\newcommand{\getnodetitle}[1]{\csname #1-title\endcsname}
\newcommand{\getnodestamps}[1]{\csname #1-stamps\endcsname}

\def\subnode#1{%
  \addcontentsline{toc}{section}{#1}%
  \refstepcounter{equation}{\bf\large\theequation.~#1}%
  \nopagebreak}

\newcommand{\note}[1][]{%
  \ifthenelse{\equal{#1}{}}{}{\addcontentsline{toc}{subsection}{#1}}%
  \notebeforesep\noindent\refstepcounter{equation}%
  \ifthenelse{\equal{#1}{}}%
  {{\S\theequation.}~}%
  {{\S\theequation.}~{\it #1:~~}}%
  \ignorespaces}

\newcommand{\nodeproclaim}[2][]{%
  \medskip\noindent\refstepcounter{equation}%
  \ifthenelse{\equal{#1}{}}%
  {{\bf #2~\theequation.~}}%
  {\addcontentsline{toc}{subsection}{#1}%
  {\bf #2~\theequation}~{\it (#1).~}}}

\newcommand{\nodeproof}{\smallskip\noindent{\bf Proof.}~}

\newcommand{\notetarget}[1]{\hypertarget{#1}\expandafter\label{#1}}

\newcommand{\notelink}[2]{\expandafter\hyperlink{#1}{#2}~\S\ref{#1}}

{%
  
  \begin{compactenum}%
  \setcounter{enumi}{\value{equation}}%
}{%
  \end{compactenum}%
}

{%
  
  \begin{compactenum}%
  \setcounter{enumii}{\value{equation}}%
}{%
  \setcounter{enumi}{\value{enumii}}\end{compactenum}%
}

\renewcommand{\notelink}[2]{\expandafter\hyperlink{#1}{#2}}
\renewcommand{\note}[1][]{\notebeforesep}

\newcommand\clD{{\mathcal D}}

\newcommand\clG{{\mathcal G}}
\newcommand\clH{{\mathcal H}}

\newcommand\clK{{\mathcal K}}

\newcommand\clM{{\mathcal M}}

\newcommand\clP{{\mathcal P}}
\newcommand\clQ{{\mathcal Q}}

\newcommand\bbR{{\mathbb R}}

\newcommand\fkg{{\mathfrak g}}

\renewcommand\![1]{{\bm{{#1}}}}
\newcommand{\qed}{$\blacksquare$}
\newcommand\onm\operatorname

\newcommand{\bigset}[2]{\bigl\{\mskip1mu #1\bigm|#2\mskip1mu\bigr\}}

\newcommand{\sset}[1]{\{\mskip1.25mu#1\mskip1.25mu\}}
\newcommand{\bigsset}[1]{\bigl\{\mskip1.25mu#1\mskip1.25mu\bigr\}}

\newcommand{\tint}{}

\newcommand{\LclH}{\mathord{\tint L_{\mskip-1mu\clH}}}
\newcommand{\LclM}{\mathord{\tint L_{\mskip-3mu\clM}}}
\newcommand{\SE}[1]{{\tint\mathit{SE}(#1)}}
\newcommand{\SO}[1]{{\tint\mathit{SO}(#1)}}
\newcommand{\TclD}{\mathord{\tint T\mskip.5mu\clD}}
\newcommand{\TclM}{\mathord{\tint T\mskip-2.25mu\clM}}
\newcommand{\TsclM}{\mathord{\tint T_{\mskip-1.5mus}\mskip-.25mu\clM}}
\newcommand{\TclQ}{\mathord{\tint T\mskip-1.mu\clQ}}

\newcommand{\TtauclQ}{\mathord{\tint T\mskip-.125mu\tauclQ}}

\newcommand{\YE}{\mathord{\tint Y_{\mskip-2muE}}}

\newcommand{\deltaA}{\mathord{\tint\delta\mskip-1muA}}
\newcommand{\deltaOmega}{\mathord{\tint\delta\mskip0mu\Omega}}
\newcommand{\deltaa}{\mathord{\tint\delta\mskip0mua}}
\newcommand{\deltadots}{\mathord{\tint\delta\mskip-.5mu\dot s}}
\newcommand{\deltadotx}{\mathord{\tint\delta\mskip-.125mu\dot x}}
\newcommand{\deltag}{\mathord{\tint\delta\mskip-.5mug}}
\newcommand{\deltas}{\mathord{\tint\delta\mskip-.5mus}}
\newcommand{\deltav}{\mathord{\tint\delta\mskip-.125mux}}
\newcommand{\deltaxi}{\mathord{\tint\delta\mskip0mu\xi}}
\newcommand{\deltax}{\mathord{\tint\delta\mskip-.125mux}}

\newcommand{\nclH}{\mathord{\tint n_{\mskip-1mu\clH}}}
\newcommand{\nclM}{\mathord{\tint n_{\mskip-3mu\clM}}}
\newcommand{\omegaL}{\mathord{\tint\omega_{\mskip-.75muL}}}
\newcommand{\se}[1]{{\tint\mathit{se}(#1)}}
\newcommand{\tauclQ}{\mathord{\tint\tau_{\mskip-1.75mu\clQ}}}

\definecolor{refcolor}{RGB}{42,93,176}
\hypersetup{colorlinks=true,linkcolor=refcolor,citecolor=refcolor}

\titleformat*{\section}{\large\bfseries}
\titlespacing\section{0pt}{10pt plus 2pt minus 1pt}{2pt plus 0pt minus 0pt}
\titleformat*{\subsection}{\bfseries}
\titlespacing\subsection{0pt}{8pt plus 2pt minus 1pt}{1pt plus 0pt minus 0pt}

\begin{document}\allowdisplaybreaks\mathtoolsset{showonlyrefs}

\thispagestyle{empty}

\begin{raggedright}
{\LARGE\bf
  The vector field of a rolling rigid body
}
\\[15pt]{
  George W.\ Patrick\\
  Department of Mathematics and Statistics,\\
  University of Saskatchewan, S7N5E6\\
  \today
}
\\[10pt]
{\color{lightgray}\rule{\textwidth}{.5pt}}
\parbox{\textwidth}{
  \vspace*{8pt}{\noindent\bf Abstract.\\}
  Nonholonomic systems are variational models commonly used for mechanical
  systems with ideal no-slip constraints. This note provides a
  differential-geometric derivation of the nonholonomic equations of motion for
  an arbitrary rigid body rolling on an arbitrary surface, via the
  semi-symplectic formalism, and in terms of shape operators (a.k.a. Weingarten
  maps). By a semi-symplectic reduction, the well-known differential equations
  in the case where the surface is a horizontal plane are shown to be
  semi-symplectic.
}
\\[9pt]
{\color{lightgray}\rule{\textwidth}{.5pt}}\\[6pt]
\end{raggedright}

\iftoggle{node}{\nodetableofcontents\bigskip}\relax

\note
\iftoggle{vector-field-to-arxiv}{\noindent}\relax
Given a configuration manifold $\clQ$, a smooth Lagrangian
$L\colon \TclQ\rightarrow\bbR$, and a (generally non-involutive) distribution
$\clD$ on $\clQ$, the evolutions $q(t)$, $t\in[a,b]$ in the
\emph{Lagrange d'Alembert model} are defined by the (fixed endpoint)
variational problem
\begin{equation}
  \mbox{$\displaystyle dS\bigl(q(t)\bigr)\cdot\delta q(t)=0$
  for all $\displaystyle\delta q(t)\in\clD$,
  with the additional constraint $\displaystyle q'(t)\in\clD$,}
\end{equation}
where the action $S$ is defined by ($q'(t)$~denotes the geometric derivative in
$\TclQ$, i.e., including the base point)
\begin{equation}
  S\bigl(q(t)\bigr)=\int_a^bL\bigl(q'(t)\bigr)\,dt.
\end{equation}
The model is called \emph{holonomic} if~$\clD$ is integrable, and otherwise it
is \emph{nonholonomic}. The usual energy is conserved, although the usual
symmetry-associated momentum may not~be. For further information on such
models, see
\cite{%
  BatesL-SniatyckiJ-1993-1,%
  MarleCM-1995-1,%
  MarleCM-1998-1,%
  SniatyckiJ-1998-1,%
  PatrickGW-2007-1%
}, and the references therein.

\note
\iftoggle{node}{
A single rigid body rolling on a stationary surface is an example of a
symmetric nonholonomic system as described above.}  { Given natural regularity
conditions on~$L$, the critical curved of the Lagrange d'Alembert model
correspond to (projections to $\clQ$) of the integral curves of a vector
field~$Y_E$ on the phase space $\clD$ itself. If $\clD=\TclQ$ then the
differential equations defined by $Y_E$ are the Euler-Lagrange equations, and
otherwise they are the Lagrange-d'Alembert equations. This note provides a
derivation of~$Y_E$ for a single rigid body rolling on a surface in Euclidean
space; see~\eqref{eq:vector-field-general} and~\eqref{eq:vector-field-reduced}.

\note
The derivation here uses the semi-symplectic formalism, where $\YE$ is
determined by the system energy and a nondegenerate antisymmetric two from on
a \emph{distribution} of the relevant phase space. If the body rolls on a
horizontal plane then the semi-symplectic system admits an $\SE2$ symmetry and
a semi-symplectic reduction shows that the well-known differential equations of
this planar system are semi-symplectic.
}\relax

\section{Lagrangian formalism}
\iftoggle{node}{
\bigskip
%%%%%%%%%%%%%%%%%%%%%%%%%%%%%%%%%%%%%%%%%%%%%%%%%%%%%%%%%%%%%%%%%%%%%%%%%%%%%%%\
%%%%%%%%%%%%%%%%%%%%%%%%%%%%%%%%%%%%%%%%%%%%%%%%%%%%%%%%%%%%%%%%%%%%%%%%%%%%%%%\
%                                                                               
\subnode{As a lagrangian system}
%                                                                               
%%%%%%%%%%%%%%%%%%%%%%%%%%%%%%%%%%%%%%%%%%%%%%%%%%%%%%%%%%%%%%%%%%%%%%%%%%%%%%%\
%%%%%%%%%%%%%%%%%%%%%%%%%%%%%%%%%%%%%%%%%%%%%%%%%%%%%%%%%%%%%%%%%%%%%%%%%%%%%%%\
}\relax

\note[Lagrangian and configuration space]
\iftoggle{vector-field-to-arxiv}{\noindent}\relax
Assume a reference body with moments of inertia~$I$ and center of mass at the
origin of the reference frame. Let the surface of the body in the reference
frame be the 2-submanifold $\clM$ and let $\clH\subseteq\bbR^3$ be the
2-submanifold on which the body rolls. Configuration of the body may be
determined by elements $(A,a,s)\in\SE3\times\clM$, with interpretation that a
point~$X$ in the reference body is located at~$AX+a$, and that the body
contacts the surface at~$As+a$. Left translating, $\Omega=A^{-1}\dot A$ and
$v=A^{-1}\dot a$, and the Lagrangian on~$T(\SE3\times\clM)$ is the left
invariant
\begin{equation}\label{eq:lagrangian}
   L=\frac12\Omega^tI\,\Omega+\frac12m|v|^2-mga\cdot\!k.
\end{equation}
Establishing contact of the the body with the fixed surface means imposing
constraints. First, the (holonomic) constraint $As+a\in\clH$ imposes that the
contact point lies on the surface. Second, assuming~$\nclM$ and~$\nclH$ are
respectively smooth choices of unit normal for~$\clM$ and $\clH$ (so both
surfaces are assumed orientable), the (holonomic) constraint
$A\,\nclM(s)=\nclH(As+a)$ imposes that the surfaces do not infinitesimally
interpenetrate at the contact point (global body-surface interpenetration
issues are not considered here). Defining~$x=As+a$, and replacing~$a$, the
configuration space is
\begin{equation}
  \clQ=\bigset{(A,s,x)\in\SO3\times\clM\times\clH}{A\,\nclM(s)=\nclH(x)}.
\end{equation}
$\clQ$ and $\TclQ$ are inserted into $\SE3\times\clM$ and $T(\SE3\times\clM)$
by the equations
\begin{equation}\label{eq:insertion-to-extended-phase-space}
  a=x-As,
  \qquad
  v=A^{-1}\dot a=A^{-1}(\dot x-\dot As-A\dot s)
  =A^{-1}\dot x-\dot s-\Omega\times s,
\end{equation}
and the Lagrangian is the pullback of~\eqref{eq:lagrangian} by these, i.e., the
result of substitution. It should be noted that the constraint
$A\,\nclM(s)=\nclH(x)$ implies physical meaning to the choice of the normals
for $\clM$ and~$\clH$. For example, if~$\clH$ is the plane~$z=0$,
and~$\nclH=-\!k$, then the choice of the \emph{outward~normal} for~$\clM$
places the body \emph{above the plane}, whereas the choice $\nclH=\!k$ places
it \emph{below}.

\note[Notations]
Recall that the Weingarten map of~$\clM$ is the vector bundle map (over the
identity) $\LclM\colon \TclM\rightarrow \TclM$ defined by
\begin{equation}
  \LclM\,\frac{ds}{dt}=-\frac d{dt}\nclM\bigl(s(t)\bigr),
\end{equation}
where~$s(t)$ is a smooth curve in~$\clM$. Similarly, $\LclH$ denotes the
Weingarten map of $\clH$.  If~$x\in\bbR^3$ then $x^\wedge$ denotes the $3\times
3$~matrix such that $x^\wedge y=x\times y$ for all~$y\in\bbR^3$.

\iftoggle{vector-field-to-arxiv}{\bigskip}\relax

\nodeproclaim[The configuration space is a manifold]{Lemma}{\sl
$\clQ$ is a $5$~dimensional submanifold of~$\SO3\times\clM\times\clH$
with tangent bundle
\begin{equation}
  \TclQ=
  \bigset{
  \bigl(\,(A,s,x),\,(\Omega,\deltas,\deltax)\,\bigr)
  }{
  (A,s,x)\in\clQ,\;
  \Omega\times\nclM(s)=\LclM(s)\,\deltas-A^{-1}\LclH(x)\,\deltax
  },
\end{equation}
and the projection~$(A,s,x)\mapsto(s,x)$ is a (trivial) principle
$\SO2$-bundle.}

\nodeproof
The map $\clQ\rightarrow S^2$ defined by $(A,s,x)\mapsto A\,\nclM(s)$ has
derivative (use left translation on the factor~$\SO3$)
\begin{equation}\label{pr:configuration-space-manifold}
  \bigl(\,(A,s,x),\,(\Omega,\deltas,\deltax)\,\bigr)
  \mapsto
  \frac d{dt}\biggr|_{t=0}
  A\onm{exp}\bigl(t\Omega^\wedge)\,\nclM(s(t)\bigr)
  =
  A\bigl(\Omega\times\nclM(s)-\LclM(s)\,\deltas\bigr)
\end{equation}
This is a submersion: take $\deltas=0$ and then
$\Omega\mapsto\Omega\times\nclM(s)$ is clearly onto the orthogonal
complement of $\nclM(s)$, i.e., onto~$T_sS^2$. So, 
$(A,s,x)\mapsto A\,\nclM(s)$ and $(A,s,x)\mapsto\nclH(x)$ are transversal, and
similarly differentiating the second of these, $\clQ$ is smooth with tangent
space at~$(A,s,x)$ the solutions~$(\Omega,\deltas,\deltax)$ to
\begin{equation}
  A\bigl(\Omega\times\nclM(s)-\LclM(s)\,\deltas\bigr)=-\LclH(x)\,\deltax,
\end{equation}
and $\onm{dim}\clQ=\onm{dim}(\SO3\times\clM\times\clH)-\onm{dim}
S^2=(3+2+2)-2=5$. If $\theta\in\bbR$ then
$\theta\cdot(A,s,x)=(A\onm{exp}(-\theta\nclM(s)^\wedge),s,x)$ defines a
right action of~$\SO2$ which, for fixed~$s$ and $x$, is free and transitive on
the~$A\in\SO3$ such that~$A\,\nclM(s)=\nclH(x)$; the assignment of the
identity of~$\SO3$ to each~$(s,x)$ is a global section.~\qed

\iftoggle{vector-field-to-arxiv}{\bigskip}\relax

\note[Rolling constraint]
The holonomic Lagrangian~\eqref{eq:lagrangian} is not regular because it does
not involve~$s$\,---\,there is no interaction of the body and the surface. To
include that interaction, impose the rolling constraint that the point on the
body at $As+a$ is instantaneously at rest,~i.e.,
\begin{equation}
  \frac d{dt}\bigl(A(t)s+a(t)\bigr)=\dot A s+\dot a=0.
\end{equation}
Ideal rolling without slipping means zero velocity of the physical location of
the fixed point on the body (at $s$) in the (inertial) frame of the surface, so
$s$ is not differentiated here. Converting to the variable~$x$, $\dot a=\dot
x-\dot As-A\dot s=-\dot As$, and the rolling constraint becomes $\dot x-A\dot
s=0$.

\note[Summary, as a lagrangian system]
\iftoggle{node}{The}{Summarizing: the} nonholonomic system for a body with surface $\clM$ rolling on a surface $\clH$
is the lagrangian system
\begin{equation}\begin{split}\label{eq:rolling-body-lagrangian-system}
  &\clQ=\bigset{(A,s,x)\in\SO3\times\clM\times\clH}{A\,\nclM(s)=\nclH(x)},
  \\
  &\TclQ=\bigset{\bigl(\,(A,s,x),\,(\Omega,\deltas,\deltax)\,\bigr)}
  {(A,s,x)\in\clQ,\;
  \Omega\times\nclM(s)=\LclM(s)\,\deltas-A^{-1}\LclH(x)\,\deltax},
  \\
  &\clD=\bigset{\bigl(\,(A,s,x),\,(\Omega,\deltas,\deltax)\,\bigr)
  \in \TclQ}{\deltax=A\,\deltas},
  \\
  &L=\frac12\Omega^tI\,\Omega+\frac12m|v|^2-mga\cdot\!k,
  \quad
  a=x-As,
  \quad
  v=A^{-1}\,\dot x-\Omega\times s-\dot s.
\end{split}\end{equation}

\section{Semi-symplectic derivation of the vector field}
\iftoggle{node}{
\bigskip
%%%%%%%%%%%%%%%%%%%%%%%%%%%%%%%%%%%%%%%%%%%%%%%%%%%%%%%%%%%%%%%%%%%%%%%%%%%%%%%\
%%%%%%%%%%%%%%%%%%%%%%%%%%%%%%%%%%%%%%%%%%%%%%%%%%%%%%%%%%%%%%%%%%%%%%%%%%%%%%%\
%                                                                               
\subnode{Vector field}\medskip
%                                                                               
%%%%%%%%%%%%%%%%%%%%%%%%%%%%%%%%%%%%%%%%%%%%%%%%%%%%%%%%%%%%%%%%%%%%%%%%%%%%%%%\
%%%%%%%%%%%%%%%%%%%%%%%%%%%%%%%%%%%%%%%%%%%%%%%%%%%%%%%%%%%%%%%%%%%%%%%%%%%%%%%\
}\relax

\note[Generic semi-symplectic summary]
\iftoggle{vector-field-to-arxiv}{\noindent}\relax
Lagrange-d'Alembert models have an equivalent semi-symplectic
formalism~(\cite{BatesL-SniatyckiJ-1993-1,SniatyckiJ-1998-1,PatrickGW-2007-1}): Given $\clQ$ and $\clD$,
a lagrangian $L\colon\clQ\rightarrow\bbR$ is called $\clD$-regular if its
second fiber derivative is nonsingular when restricted to~$\clD$. The
distribution $\clK_\clD\equiv\TclD\cap (\TtauclQ)^{-1}\clD$, where
$\tauclQ\colon \TclQ\rightarrow\clQ$ is the projection, has fiber dimension
twice that of~$\clD$, and $\omegaL$~is nonsingular on~$\clK_\clD$ if and only
if~$L$ is $\clD$-regular, in which case
\begin{equation}\label{eq:generic-semisymplectic-vector-field}
  \mbox{$\displaystyle (i_{\YE}\omegaL-dE)\bigl|\clK_\clD=0$,
  where $\displaystyle \YE(\clQ)\subseteq\clK_D$,}
\end{equation}
defines a vector field~$\YE$ with integral curves exactly the solutions of the
Lagrange-d'Alembert variational principle. In general one is led to a category
defined by a nondegenerate antisymetric two form with domain a (generally
nonintegrable) distribution. The semi-symplectic formulation is advantageous
because it has this formula for the evolution vector field\,---\,the
Lagrange-d'Alembert equations have already been geometrically determined
as~\eqref{eq:generic-semisymplectic-vector-field}.

\note[Regularity of the lagrangian]
Since~$L$ is fiberwise bilinear, regularity is equivalent to~$\Omega=0$, 
$\dot s=0$, and~$\dot x=0$ whenever~$\Omega^tI\,\Omega+m|v|^2=0$.  Assuming $I$
is positive definite and~$m>0$, the latter is equivalent to~$\Omega=0$
and~$v=0$, i.e., $\dot x=A\dot s$ (within $\TclQ$). Restricting to~$\TclQ$
leads to $\LclM(s)\dot s-A^{-1}\LclH(x)\,A\dot s=0$, so $L$ is $\clD$-regular
if and only
\begin{equation}
  \Lambda_{A,s,x}\colon \TsclM\to\TsclM,
  \qquad
  \Lambda_{A,s,x}\equiv\LclM(s)-A^{-1}\LclH(x)\,A
\end{equation}
is fiberwise nonsingular for all~$(A,s,x)\in\clQ$.

\note[As a semi-symplectic system]
Another advantage of the semi-symplectic formalism is an early clear emphasis
and identification of the relevant phase space~$\clD$, which by
Lemma~\ref{pr:configuration-space-manifold} is the subset of~$\TclQ$ satisfying
$\dot x=A\dot s$ and $\Lambda_{A,s,x}\dot s=\Omega\times\nclM(s)$, and which,
if $L$ is regular, may be identified with~$\clP=\clQ\times\bbR^3$ by
\begin{equation}\label{eq:vector-field-known-part}
  (A,s,x,\Omega)\leftrightarrow
  \bigl(\,
  (A,s,x),\,
  (\Omega,\Lambda_{A,s,x}^{-1}(\Omega\times\nclM(s)),
  A\Lambda_{A,s,x}^{-1}(\Omega\times\nclM(s))
  \,\bigr).
\end{equation}
So from the outset one seeks differential equations for $dA/dt$, $ds/dt$,
$dx/dt$, and~$d\Omega/dt$, which is not entirely obvious apriori because from
the variational principle one might have anticipated second order differential
equations for $s$ or~$x$. Since every evolution has derivative in~$\clD$ and is
second order, three of the required differential equations are known:
\begin{equation}
  \frac{dA}{dt}=A^{-1}\Omega,
  \qquad
  \Lambda_{A,s,x}\frac{ds}{dt}=\Omega\times\nclM,
  \qquad
  \frac{dx}{dt}=A\,\frac{ds}{dt}.
\end{equation}
Only the differential equation for $d\Omega/dt$ need be determined.

\note
To identify the rolling body system as semi-symplectic, assuming regularity, it
is required to find on $\clP$ the distribution~$\clK_\clP$, the Lagrange
two-form~$\omegaL$, and the pullback of the energy~$E$, all of which are
defined by pullback to~$\clP$.

\note[Semi-symplectic distribution]\notetarget{nt:K-P-distibution}
Using left translation with the first factor~$\SO3$ of~$\clP$,
the pullback of~$\clK_\clD$ to the distribution on $\clK_\clP$ on $\clP$ is
the pullback of $\clD$ by the projection $(A,s,x,\Omega)\mapsto(A,s,x)$, i.e.
\begin{equation}\label{eq:K-P-distibution}
  \clK_\clP=
  \bigset{
  \bigl(\,(A,s,x,\Omega),\,(\deltaA,\deltas,\deltax,\deltaOmega)\,\bigr)
  }{
  (A,s,x)\in\clQ,
  \;
  \deltaA\times\nclM(s)=\Lambda_{(a,s,x)}\,\deltas,
  \;\deltas=A\,\deltax}.
\end{equation}
This may be viewed as determining~$\deltas$ and~$\deltax$ with free and
uncoupled $\deltaA$ and $\deltaOmega$, and hence has fiber dimension~$6$.
Analogously, in~\eqref{eq:rolling-body-lagrangian-system}, $\deltas$ and
$\deltax$ are determined from a free $\Omega$, so the fiber dimension of $\clD$
is~$3$.

\note
It is an \marginpar{\dbend} error to substitute the constraint distribution
into the Lagrangian \emph{before} calculating the Lagrange one form. This is
the point in the semi-symplectic formalism which avoids obtaining incorrect
evolution equations by substituting the constraint into the Lagrangian
before varying the action.

\note[Semi-symplectic two-form]
The Lagrange forms are natural with respect to lifts of diffeomorphisms, so it
suffices to pull back~$\omegaL$ defined by~\eqref{eq:lagrangian} as a function
on~$T(\SE3)\times\bbR^3$, and \eqref{eq:lagrangian} is independent of $s$, so a
formula for $\omegaL$ with $L$ regarded as a left invariant Lagrangian of
$T(\SE3)$ will do. The general formula for the Lagrange two-form of a left
invariant Lagrangian~$L(\xi)$ on a Lie group~$G=\sset{g}$, where
$\fkg=\sset{\xi}$ is the Lie algebra and $\xi\in\fkg$, is
\begin{equation}
  \omegaL(g,\xi)
  \bigl(\,
  (g,\xi,\deltag_1,\deltaxi_1),\,
  (g,\xi,\deltag_2,\deltaxi_2)
  \,\bigr)
  =
  D^2L(\xi)\,(\deltaxi_2,\deltag_1)
  -D^2L(\xi)\,(\deltaxi_1,\deltag_2)
  +DL(\xi)\,\relax[\deltag_1,\deltag_2],
\end{equation}
and since Lie bracket of $\se3=\sset{(\xi,u)}$ is
\begin{equation}
  [(\xi,u),(\eta,v)]=(\xi\times\eta,\xi\times v-\eta\times u),
\end{equation}
the Lagrange two-form of $L$ on $T(\SE3)\times\clM$ is
\begin{equation}\begin{split}\label{eq:omega-L-extended-phase-space}
  &\omegaL(A,s,\Omega,v)
  \bigl(\,
  (\deltaA_1,\deltaa_1,\deltaOmega_1,\deltav_1),\,
  (\deltaA_2,\deltaa_2,\deltaOmega_2,\deltav_2)
  \,\bigr)
  \\
  &\qquad\mbox{}
  =\bigl(
  (I\,\deltaOmega_2)\cdot\deltaA_1+m\,\deltav_2\cdot\deltaa_1
  \bigr)
  -\bigl(
  (I\,\deltaOmega_1)\cdot\deltaA_2+m\,\deltav_1\cdot\deltaa_2
  \bigr)
  \\
  &\qquad\qquad\mbox{}
  +(I\Omega)\cdot(\deltaA_1\times\deltaA_2)
  +mv\cdot(\deltaA_1\times\deltaa_2-\deltaA _2\times\deltaa_1).
\end{split}\end{equation}
Obtaining the pullback of~\eqref{eq:omega-L-extended-phase-space} to~$\clD$
means substituting the derivatives of
\eqref{eq:insertion-to-extended-phase-space}, i.e.,
\begin{equation}\begin{split}\label{eq:pullback-1}
  \begin{array}{ll}
  \displaystyle
  a=x-As
  \qquad&\displaystyle
  \deltaa=\deltax-A\,(\deltaA\times s)-A\,\deltas,
  \\
  \displaystyle
  v=A^{-1}\dot x-\dot s-\Omega\times s,
  \qquad&\displaystyle
  \deltav=
  -\deltaA\times(A^{-1}\dot x)+A^{-1}\deltadotx-\deltadots
  -\deltaOmega\times s-\Omega\times\deltas.
  \end{array}
\end{split}\end{equation}
Here it is useful to realize that the semi-symplectic form is only required
on~$\clK_\clD$ and may be replaced by any two-form with equal values
on that. Since~$\clK_\clD$ is defined by
\begin{equation}
  \dot x = A\dot s,
  \qquad
  \deltax=A\,\deltas,
  \qquad
  \deltadotx=A\,(\deltaA\times\dot s)+A\,\deltadots,
\end{equation}
these may be substituted into~\eqref{eq:pullback-1} to obtain the
simpler
\begin{equation}
  \deltaa=-\deltaA\times s,
  \qquad
  v=-\Omega\times s,
  \qquad
  \deltav=-\deltaOmega\times s-\Omega\times\deltas.
\end{equation}
with the result (the symbol~$\cong$ means equal on~$\clK_D$)
\begin{equation}\begin{split}\label{eq:semisymplectic-form}
  \omegaL
  &\cong
  (I\,\deltaOmega_2)\cdot\deltaA_1
  +m(-\deltaOmega_2\times s-\Omega\times\deltas_2)
  \cdot(-\deltaA_1\times s)
  \\
  &\qquad
  \mbox{}-(I\,\deltaOmega_1)\cdot\deltaA_2
  -m(-\deltaOmega_1\times s-\Omega\times\deltas_1)
  \cdot(-\deltaA_2\times s)
  \\
  &\qquad
  \mbox{}+(I\Omega)\cdot(\deltaA_1\times\deltaA_2)
  -m(\Omega\times s)\cdot
  \bigl(
  \deltaA_1\times(-\deltaA_2\times s)
  -\deltaA _2\times(-\deltaA_1\times s)
  \bigr)
  \\
  &=
  \bigl(
  I\,\deltaA_1
  -ms\times(s\times\deltaA_1)
  \bigr)
  \cdot\deltaOmega_2
  -m(\Omega\times\deltas_2)\cdot(s\times\deltaA_1)
  \\&\qquad
  \mbox{}-
  \bigl(
  I\,\deltaOmega_1
  -ms\times(s\times \deltaOmega_1)
  -m(\Omega\times\deltas_1)\cdot(s\times\deltaA_2
  \bigr)
  \cdot\deltaA_2
  \\&\qquad
  \mbox{}+(I\Omega)\cdot(\deltaA_1\times\deltaA_2)
  -m(\Omega\times s)\cdot
  \bigl(
  s\times(\deltaA_1\times\deltaA_2)
  \bigr)
  \\
  &=
  (\tilde I\,\deltaA_1)\cdot\deltaOmega_2
  -(\tilde I\,\deltaA_2)\cdot\deltaOmega_1
  +(\tilde I\Omega)\cdot(\deltaA_1\times\deltaA_2)
  \\&\qquad
  \mbox{}+m(\Omega\times\deltas_1)\cdot(s\times\deltaA_2)
  -m(\Omega\times\deltas_2)\cdot(s\times\deltaA_1).
\end{split}\end{equation}
To view this as a two-form restricted to $\clK_\clP$, regard $\deltaA$
and~$\deltaOmega$ as free and restrict $\deltas$ and $\deltax$ as
in~\eqref{eq:K-P-distibution}, assuming of course that
$A\,\nclM(s)=\nclH(x)$.
In the same way, one requires the energy $E$ only restricted to $\clD$, so
\begin{equation}
  E\cong\frac12\Omega^tI\,\Omega+\frac12m|\Omega\times s|^2+mg(x-As)\cdot\!k
  =\frac12\Omega^t\tilde I\,\Omega+mg(x-As)\cdot\!k,
\end{equation}
and,
\begin{equation}\begin{split}\label{eq:semisymplectic-energy-derivative}
  dE&\cong(I\Omega)\cdot\deltaOmega_2
  +mv\cdot\deltav
  -mgA\,(\deltaA_2\times s)\cdot\!k
  \\
  &=
  (I\Omega)\cdot\deltaOmega_2
  +m(\Omega\times s)\cdot(\deltaOmega_2\times s+\Omega\times\deltas_2)
  -mg(s\times A^{-1}\!k)\cdot\deltaA_2
  \\
  &=
  (\tilde I\Omega)\cdot\deltaOmega_2
  -m(s\times\Omega)\cdot(\Omega\times\deltas_2)
  -mg(s\times A^{-1}\!k)\cdot\deltaA_2.
\end{split}\end{equation}

\note[Vector field computation actual]
To find the vector field $\YE$ replace $1$-subscripted quantities such as
$\deltaOmega_1$ with their corresponding derivatives $d\Omega/dt$, and set
\eqref{eq:semisymplectic-form} to the
derivative~\eqref{eq:semisymplectic-energy-derivative} of~$E$, for all
$(\deltaA_2,\deltas_2,\deltax_2,\deltaOmega_2)\in\clK_\clP$. As already
\notelink{nt:K-P-distibution}{noted}, in this context $\deltaOmega_2$ and
$\deltaA_2$ are free and uncoupled, and $\deltas_2=0$ if $\deltaA_2=0$. Set
$\deltaA_2=0$ and $\deltas_2=0$ to obtain $\tilde I\,\deltaA_1=\tilde I\Omega$
i.e. $A^{-1}dA/dt=\Omega$, which is already
known~\eqref{eq:vector-field-known-part}. Substituting back, all the
$\deltas_2$ cancel,
\begin{equation}\begin{split}
  &-\bigl((\tilde I\,\deltaA_2)\cdot\deltaOmega_1
  +(\tilde I\Omega)\cdot(\Omega\times\deltaA_2)
  +m(\Omega\times\deltas_1)\cdot(s\times\deltaA_2)\\
  &\qquad=
  -\bigl(\tilde I\,\deltaOmega_1
  +(\tilde I\Omega)\times\Omega
  +m(\Omega\times\deltas_1)\times s\bigr)\cdot\deltaA_2\\
  &\qquad=
  -mg(s\times A^{-1}\!k)\cdot\deltaA_2,
\end{split}\end{equation}
and finally
\begin{equation}
\tilde I\frac{d\Omega}{dt}=(\tilde I\Omega)\times\Omega
+ms\times\left(\frac{ds}{dt}\times\Omega\right)+mgs\times(A^{-1}\!k).
\end{equation}

\iftoggle{node}{
\note[Summary, of the vector field and dynamical system] The}
{Summarizing: the} dynamical system
corresponding to the Lagrange-d'Alembert variational
principle~\eqref{eq:rolling-body-lagrangian-system} are
\begin{equation}\begin{split}\label{eq:vector-field-general}
  &\clP=\bigset{(A,s,x,\Omega)\in\SO3\times\clM\times\clH\times\bbR^3}{A\nclM(s)=\nclH(x)},
  \quad
  E=\frac12\Omega^t\tilde I\,\Omega+mg(x-As)\cdot\!k,
  \\
  &\Lambda_{A,s,x}\,\frac{ds}{dt}=\Omega\times\nclM(s),
  \quad
  \tilde I\,\frac{d\Omega}{dt}
  =(\tilde I\Omega)\times\Omega
  +ms\times\left(\frac{ds}{dt}\times\Omega+g(A^{-1}\!k)\right),
  \quad
  A^{-1}\frac{dA}{dt}=\Omega^\wedge,
  \\
  &\frac{dx}{dt}=A\,\frac{ds}{dt},
  \quad
  \Lambda_{A,s,x}=\LclM(s)-A^{-1}\LclH(x)\,A,
  \quad
  \tilde I=I-m(s^\wedge)^2.
\end{split}\end{equation}

\iftoggle{vector-field-to-arxiv}{\vspace*{-5pt}}\relax

\section{Rolling on a horizontal plane; semi-symplectic reduction}
\iftoggle{node}{
\bigskip
%%%%%%%%%%%%%%%%%%%%%%%%%%%%%%%%%%%%%%%%%%%%%%%%%%%%%%%%%%%%%%%%%%%%%%%%%%%%%%%\
%%%%%%%%%%%%%%%%%%%%%%%%%%%%%%%%%%%%%%%%%%%%%%%%%%%%%%%%%%%%%%%%%%%%%%%%%%%%%%%\
%                                                                               
\subnode{The special case of rolling on a plane}
%                                                                               
%%%%%%%%%%%%%%%%%%%%%%%%%%%%%%%%%%%%%%%%%%%%%%%%%%%%%%%%%%%%%%%%%%%%%%%%%%%%%%%\
%%%%%%%%%%%%%%%%%%%%%%%%%%%%%%%%%%%%%%%%%%%%%%%%%%%%%%%%%%%%%%%%%%%%%%%%%%%%%%%\
}\relax

\note[Specialization to the planar case]
\iftoggle{vector-field-to-arxiv}{\noindent}\relax
This is the special case where~$\clH$ is the $x,y$ plane, $\LclH=0$
and~$\nclH=-\!k$ (so that the body is above the plane when~$\nclM$ is the
outward normal). The group
\begin{equation}
  \clG\equiv\bigset{(B,b)\in\SE3}{B\!k=\!k,b\cdot\!k=0}\cong\SE2
\end{equation}
acts on the semisymplectic
phase space~$\clP$ by 
\begin{equation}\label{eq:P-action}
  (B,b)\bigl(A,s,x,\Omega)=(BA,s,Bx+b,\Omega)
\end{equation}
and the projection to~$\clM\times\bbR^3=\sset{(s,\Omega)}$ is a quotient
map. The vector field~\eqref{eq:vector-field-general} is equivariant and the
differential equations for~$s$ and~$\Omega$ close: on $\clP$,
$A\,\nclM(s)=\nclH(x)=-\!k$ so $\nclM(s)=-A^{-1}\!k$ can be substituted.  Also,
the energy drops to the quotient by using 
$\!k\cdot(x-As)=-\!k\cdot As=\nclM\cdot s$, leading to the dynamical system
\begin{equation}\begin{split}\label{eq:vector-field-reduced}
  &\bar\clP=\bigsset{(s,\Omega)\in\clM\times\bbR^3},
  \quad
  E=\frac12\Omega^t\tilde I\Omega+mg\nclM\cdot s.
\\
  &\LclM\,\frac{ds}{dt}=\Omega\times\nclM(s),
  \quad
  \tilde I\,\frac{d\Omega}{dt}=(\tilde I\Omega)\times\Omega
  +ms\times\left(\frac{ds}{dt}\times\Omega-g\nclM\right),
  \quad
  \tilde I=I-m(s^\wedge)^2.
\end{split}\end{equation}
Equations~\eqref{eq:vector-field-reduced} are the same as equations (8a--c)
in~\cite{GarciaA-HubbardM-1988-1}, equations~(1.1) and~(1.2)
of~\cite{BorisovAV-MamaevIS-2002-1}, and equations~(4)
of~\cite{BorisovAV-MamaevIS-2003-1} (after replacing $d\nclM/dt=-\LclM\,ds/dt$
and accounting for the choice of unit normal).

\note[In terms of surface coordinates]
\iftoggle{vector-field-to-arxiv}{In passing, if}{Suppose}
$y^a$, $a=1,2$, are coordinates on $\clM$, so that $\clM$ is the image of an
immersion $s(y)$, arranged so that the outward normal is
\begin{equation}
  \nclM=\frac{\partial s}{\partial y^1}\times
  \frac{\partial s}{\partial y^2}
\end{equation}
Let $g_{ab}$ and~$L_{ab}$ be the first and second fundamental forms of $\clM$,
and let $L^a_b$ be the Weingarten map, so
\begin{equation}
  g_{ab}=\frac{\partial s}{\partial y^a}\cdot\frac{\partial s}{\partial y^a},
  \qquad
  L_{ab}=\nclM\cdot\frac{\partial^2 s}{\partial y^a\,\partial y^b},
  \qquad 
  L^a_b=g^{ac}L_{bc},
\end{equation}
giving a $2\times2$ matrix $L\equiv[L_{ab}]$ that depends on~$y$. From the left
side of the differential equation for~$ds/dt$
in \eqref{eq:vector-field-reduced},
\begin{equation}
  \frac{\partial s}{\partial y^a}\cdot\LclM\,\frac{ds}{dt}
  =\frac{\partial s}{\partial y^a}\cdot
  \left(\LclM\,\frac{\partial s}{\partial y^b}\right)\frac{dy^b}{dt}
  =\frac{\partial s}{\partial y^a}\cdot
  \left(L^c_b\frac{\partial s}{\partial y^c}\right)\frac{dy^b}{dt}
  =g_{ac}L^c_b\frac{dy^b}{dt}
  =L_{ab}\frac{dy^b}{dt}
  =\left[L\,\frac{dy^b}{dt}\right]_a,
\end{equation}
while on the right side,
\begin{equation}
  \frac{\partial s}{\partial y^a}\cdot(\Omega\times\nclM)
  =\biggl(\nclM\times\frac{\partial s}{\partial y^a}\biggr)\cdot\Omega
  =\left[B^t\Omega\right]_a,
  \qquad
  B=\nclM^\wedge\,\frac{\partial s}{\partial y},
\end{equation}
giving the equations of motion
\begin{equation}\label{eq:vector-field-reduced-coordinate}
  L\frac{dy}{dt}=B^t\Omega,
  \qquad
  \tilde I\,\frac{d\Omega}{dt}+\left(ms^\wedge\Omega^\wedge
  \frac{\partial s}{\partial y}\right)
  \frac{dy}{dt}
  =(\tilde I\Omega)\times\Omega+mg\nclM\times s,
\end{equation}
where $\tilde I=I-m(s^\wedge)^2$, the $2\times2$ matrix $L$, the $3\times 2$
matrix $B$, and the three-vectors $s$ and $\nclM$, are all given functions of
$y$.

\iftoggle{node}{
\note[Verify energy conservation]
To check energy conservation in~\eqref{eq:vector-field-reduced},
using the Jacobi identity,
\begin{equation}\begin{split}
  \Omega\cdot\frac{d\tilde I}{dt}\Omega
  &=
  \Omega\cdot\left(
  -m\frac{ds}{dt}\times(s\times\Omega)
  -ms\times\left(\frac{ds}{dt}\times\Omega\right)
  \right)
  \\
  &=
  \Omega\cdot\left(
  ms\times\left(\Omega\times\frac{ds}{dt}\right)
  +m\Omega\times\left(\frac{ds}{dt}\times s\right)
  -ms\times\left(\frac{ds}{dt}\times\Omega\right)
  \right)
  \\
  &=
  2m\Omega\cdot\left(
  s\times\left(
  \Omega\times\frac{ds}{dt}
  \right)
  \right),
\end{split}\end{equation}
so that
\begin{equation}\begin{split}
  \frac{dE}{dt}&=
  \Omega\cdot\tilde I\,\frac{d\Omega}{dt}
  +\frac12\Omega\cdot\frac{d\tilde I}{dt}\Omega
  +mg\left(\frac{d\nclM}{dt}\cdot s+\nclM\cdot\frac{ds}{dt}\right)
  \\
  &=\Omega\cdot
  \left(
  (\tilde I\Omega)\times\Omega
  +ms\times\left(\frac{ds}{dt}\times\Omega-g\nclM\right)
  \right)
  +m\Omega\cdot\left(
  s\times\left(
  \Omega\times\frac{ds}{dt}
  \right)
  \right)
  \\
  &\qquad\mbox{}
  +mg\left(-\LclM\frac{ds}{dt}\cdot s+\nclM\cdot\frac{ds}{dt}\right)
  \\
  &=m\Omega\cdot\left(
  s\times\left(\frac{ds}{dt}\times\Omega-g\nclM\right)
  \right)
  +m\Omega\cdot\left(
  s\times\left(
  \Omega\times\frac{ds}{dt}
  \right)
  \right)
  +mg(\nclM\times\Omega)\cdot s
  \\
  &=m\Omega\cdot\left(
  s\times\left(\frac{ds}{dt}\times\Omega\right)
  \right)
  +m\Omega\cdot\left(
  s\times\left(
  \Omega\times\frac{ds}{dt}
  \right)
  \right)
  -mg\Omega\cdot(s\times\nclM)
  +mg(\nclM\times\Omega)\cdot s
  \\
  &=0.
\end{split}\end{equation}}\relax

\note[Symmetry, as a semi-symplectic system]
The formula $(B,b)(A,s,x)=(BA,s,Bx+b)$, $(B,b)\in\clG$, is an action on $\clQ$,
because $(A,s,x)\in\clQ$ implies $(BA,s,Bx+b)\in\clQ$, since
\begin{equation}
  (BA)\,\nclM(s)=BA\,\nclM(s)=B\,\nclH(x)=B\!k=\!k=\nclH(Bx+b),
\end{equation}
The action lifts to $\TclQ$ as
\begin{equation}\begin{split}\label{eq:SE2-action-lifted}
  (B,b)\bigl(\,(A,s,x),\,(\Omega,\dot s,\dot x)\,\bigr)
  &=\left.\frac d{d\epsilon}\right|_{t=0}(B,b)(A+\epsilon A\Omega^\wedge,
  \dot s,x+\epsilon\dot x)
\\
  &=\bigl(\,(BA,s,Bx+b),\,(A^{-1}B^{-1}BA\Omega^\wedge,\dot s,B\dot x)\,\bigr)
\\
  &=\bigl(\,(BA,s,Bx+b),\,(\Omega,\dot s,B\dot x)\,\bigr).
\end{split}\end{equation}
If $\dot x=A\dot s$ then $B\dot x=B(A\dot s)=(BA)\dot s$ so
the \eqref{eq:SE2-action-lifted} preserves the rolling constraint, restricts to
an action on $\clD$, and induces on $\clP$ the action~\eqref{eq:P-action}. The
Lagrangian is invariant:
\begin{equation}\begin{split}
  &L\bigl((B,b)(\,(A,s,x),\,(\Omega,\dot s,\dot x)\,)\bigr)
\\
  &\qquad\qquad=
  L\bigl(\,(BA,s,Bx+b),\,(\Omega,\dot s,B\dot x)\,\bigr)
\\
  &\qquad\qquad=
  \frac12\Omega^tI\,\Omega+\frac12|(BA)^{-1}B\dot x-\Omega\times s-\dot s|^2
  -mg(Bx+b-BAs)\cdot\!k
\\
  &\qquad\qquad=
  \frac12\Omega^tI\,\Omega+\frac12|A^{-1}\dot x-\Omega\times s-\dot s|^2
  -mg(x-As)\cdot B^t\!k-mgb\cdot\!k
\\
  &\qquad\qquad=
  \frac12\Omega^tI\,\Omega+\frac12|A^{-1}\dot x-\Omega\times s-\dot s|^2
  -mg(x-As)\cdot\!k
\\
  &\qquad\qquad=L\bigl(\,(A,s,x),\,(\Omega,\dot s,\dot x)\,\bigr).
\end{split}\end{equation}
Consequently, $\clG$ acts symplectically with respect to $\omega_L$, and hence
acts by semi-symplectomorphisms.

\note[Momentum]
The Lie algebra of $\clG$ is $\bbR\times\bbR^2=\sset{(\xi^r,\xi^a)}$ and
the infinitesimal generator of the action is
\begin{equation}
  \left.\frac d{d\epsilon}\right|_{\epsilon=0}
  \bigl(A^{-1}(\!1+\epsilon\xi^r)
  \!k^\wedge A,s,(\!1+\epsilon\xi^r\!k^\wedge)x+\epsilon\xi^a\bigr)
  =
  \bigl(\,(A,s,x),\,(\xi^rA^{-1}\!k,0,\xi^r\!k\times x+\xi^a)\,\bigr).
\end{equation}
The momentum associated to $\xi=(\xi^r,\xi^a)$ at the state
$\bigl(\,(A,s,x),\,(\Omega,\dot s,\dot x)\,\bigr)$ is
\begin{equation}\begin{split}
  J_\xi&=\left.\frac d{d\epsilon}\right|_{\epsilon=0}
  L\bigl(\,(A,s,x),\,(\Omega+\epsilon\xi^rA^{-1}\!k,\dot s,\dot x
  +\epsilon\xi^r\!k\times x+\epsilon\xi^a)\,\bigr)
\\
  &=\Omega^tI\,(\xi^rA^{-1}\!k)+mv^t\left.\frac d{d\epsilon}\right|_{\epsilon=0}
  \Bigl(A^{-1}(\dot x+\epsilon\xi^r\!k\times x+\epsilon\xi^a)
  -(\Omega+\epsilon\xi^rA^{-1}\!k)\times s
  -\dot s\Bigr)
\\
  &=\Omega^tI\,(\xi^rA^{-1}\!k)+mv^t\Bigl(A^{-1}(\xi^r\!k\times x+\xi^a)
  -(\xi^rA^{-1}\!k)\times s\Bigr)
\\
  &=\xi^r\Bigl(AI\Omega+m(x-As)\times Av\Bigr)\cdot\!k+mAv\cdot\xi^a.
\end{split}\end{equation}
Pulling this back to $\clP$ means $\dot x=A\dot s$ and 
$\Lambda_{A,s,x}\dot s=\Omega\times\nclM(s)$, resulting in
$v=-\Omega\times s$ and
\begin{equation}\begin{split}
  J_\xi
  &=
  \xi^r\Bigl(AI\Omega-m(x-As)\times(\Omega\times s)\Bigr)\cdot\!k
  -mA(\Omega\times s)\cdot\xi^a
  \\
  &=
  \xi^r\Bigl(A\tilde I\Omega-mx\times(\Omega\times s)\Bigr)\cdot\!k
  -mA(\Omega\times s)\cdot\xi^a
\end{split}\end{equation}

\note[Summary as a symmetric semi-symplectic system] 
The phase space is $\clP$, which assuming regularity is diffeomorphic to the
original rolling distribution $\clD$. The distribution $\clK_\clD$ within
$\clP$ corresponds to the second order part of $T\clD$. From the above, the
relevant symmetric semi-symplectic formalism is
\begin{equation}\begin{split}\label{eq:semisymplectic-summary}
  &\clP=\bigset{(A,s,x,\Omega)\in\SO3\times\clM\times\clH\times\bbR^3}
  {A\,\nclM=-\!k},
\\
  &T\clP=
  \bigset{
  \bigl(\,(A,s,x,\Omega),\,(\deltaA,\deltas,\deltax,\delta\Omega)\,\bigr)
  }{
  (A,s,x)\in\clQ,
  \;
  \deltaA\times\nclM=\LclM\,\deltas},
\\
  &\clK_\clP=
  \bigset{
  \bigl(\,(A,s,x,\Omega),\,(\deltaA,\deltas,\deltax,\delta\Omega)\,\bigr)
  \in T\clP}{\deltax=A\,\deltas},
\\
  &\omega_L
  \cong
  (\tilde I\,\deltaA_1)\cdot\delta\Omega_2
  -(\tilde I\,\deltaA_2)\cdot\delta\Omega_1
  +(\tilde I\Omega)\cdot(\deltaA_1\times\deltaA_2)
\\
  &\qquad\qquad\qquad
  \mbox{}+m(\Omega\times\deltas_1)\cdot(s\times\deltaA_2)
  -m(\Omega\times\deltas_2)\cdot(s\times\deltaA_1),
\\
  &E=\frac12\Omega^t\tilde I\,\Omega+mg(x-As)\cdot\!k,
\\
  &dE\cong(\tilde I\Omega)\cdot\delta\Omega
  -m(s\times\Omega)\cdot(\Omega\times\deltas)
  +mg(s\times\nclM)\cdot\deltaA,
\\
  &\clG=\bigset{(B,b)\in\SE3}{B\!k=\!k,b\cdot\!k=0}\equiv\SE2,
\\
  &(B,b)(A,s,x,\Omega)=(BA,s,Bx+b,\Omega),
\\
  &\xi(A,s,x,\Omega)=(\xi^rA^{-1}\!k,0,\xi^r\!k\times x+\xi^a,0),
\\
  &J=\bigl(A\tilde I\Omega-mx\times(\Omega\times s)\bigr)\cdot\!(\xi_r\!k)
  -mA(\Omega\times s)\cdot\xi^a.
\end{split}\end{equation}
$\xi(A,s,x,\Omega)\in\clK_\clP$ for all $(A,s,x,\Omega)$ implies $\xi=0$ (the
$\deltas=0$ , so there is no semi-hamiltonian part of the symmetry and there
is no conserved momentum~(\cite{PatrickGW-2007-1}). Consequently, the
nonholonomic reduced phase space is
$\pi\colon\clP\rightarrow\bar\clP\equiv\clP/\clG$ with nonholonomic
distribution
$\bar\clK_{\bar\clP}=T\pi\bigl((\onm{ker}T\pi\cap\clK)^{\omega\perp}\bigr)$.
$\omega_L$ drops to a nondegerate two form on $\bar\clK_{\bar\clP}$, $E$
also drops. The resulting semi-symplectic equations must be be the same
as~\eqref{eq:vector-field-reduced}\,---\,this is verified 
\notelink{nt:verify-reduced-eq}{below}.

\note[Semi-symplectic reduction]
The map $\SO3\times\clM\times\clH\times\bbR^3\rightarrow
S^2\times\clM\times\bbR^3$ by $(A,s,x,\Omega)\mapsto(A^{-1}\!k,s,\Omega)$ is a
quotient for the action of $\clG$, because
\begin{equation}
  (B,b)(A,s,x,\Omega)=(BA,s,Bx+b,\Omega)
  \mapsto
  \bigl((BA)^{-1}\!k,s,\Omega\bigr)
  =
  \bigl(A^{-1}B^{-1}\!k,s,\Omega\bigr)
  =\bigl(A^{-1}\!k,s,\Omega\bigr),
\end{equation}
while $\pi(A,s,x,\Omega)=\pi(\tilde A,\tilde s,\tilde x,\tilde \Omega)$ implies
$A\!k=\tilde A\!k$, $s=\tilde s$, and $x=\tilde x$, from which $B=A\tilde
A^{-1}$ and $b=x=B\tilde x$ provides $(B,b)\in\clG$ such that
$(B,b)(A,s,x,\Omega)=(\tilde A,\tilde s,\tilde x,\tilde \Omega)$. The
restriction to $\clP$ has $\nclM=-A^{-1}\!k$ so
\begin{equation}
  \bar P=\clP/\clG=\clM\times\bbR^3=\sset{(s,\Omega)},
  \qquad
  \pi(A,s,x,\Omega)=(s,\Omega)
  \quad
  \mbox{(restricted to $A\,\nclM=-\!k$).}
\end{equation}
This makes 
\begin{equation}\begin{split}
  \onm{ker}(T\pi\cap\clK)
  &=
  \bigset{
  \bigl(\,(A,s,x,\Omega),\,(\deltaA,\deltas,\deltax,\delta\Omega)\,\bigr)
  \in T\clP
  }{
  \deltas=0,
  \;
  \;\delta\Omega=0,
  \;
  \deltax=A\,\deltas}
  \\
  &=
  \bigset{
  \bigl(\,(A,s,x,\Omega),\,(\deltaA,0,0,0)\,\bigr)
  }{
  \deltaA\in\bbR\,\nclM},
\end{split}\end{equation}
and required is the symplectic complement of this in $\clK$. For that, put
$\deltaA_1=\nclM$, $\deltas_1=0$, $\delta\Omega_1=0$, and
$\LclM\,\deltas_2=\deltaA_2\times\nclM$ into $\omega_L=0$,
obtaining
\begin{equation}
  (\tilde I\,\nclM)\cdot\delta\Omega_2
  -(\tilde I\Omega)\cdot(\LclM\,\deltas_2)
  -m(\Omega\times\deltas_2)\cdot(s\times\nclM)=0,
  \quad
  (\deltaA_2,\deltas_2,\deltax_2,\delta\Omega_2)\in\clK_\clP.
\end{equation}
which refers only to $\deltas_2$ and $\delta\Omega_2$. Any such can be arranged
into $\clK_\clP$, so
$\bar\clK_{\bar\clP}=T\pi\bigl((\onm{ker}T\pi\cap\clK)^{\omega\perp}\bigr)$ is
defined by
\begin{equation}
  (\tilde I\,\nclM)\cdot\delta\Omega
  -(\tilde I\Omega)^t\LclM\,\deltas
  +m(\nclM\cdot\Omega)(s\cdot\deltas)
  =0,
\end{equation}
or equivalently,
\begin{equation}\label{eq:reduced-K-constraint}
  \nclM^t\tilde I\,\delta\Omega
  -\Omega^t\tilde I\LclM\,\deltas
  +m(\nclM\cdot\Omega)s^t\,\deltas=0.
\end{equation}
To calculate $\bar\omega_L(s,\Omega)\bigl((\deltas_1,\delta\Omega_1), (\delta
s_2,\delta\Omega_2)\bigr)$ use those same $\deltas_i$ and $\delta\Omega_i$ and
substitute into the expression for $\omega_L$
in~\eqref{eq:semisymplectic-summary} any $\deltaA_i$ such that
$\deltaA_i\times\nclM=\LclM\,\deltas_i$, e.g.,
$\deltaA_i=\nclM\times \LclM\,\deltas_i$ (there is no $\deltax$ in the formula
for $\omega_L$ anyway). Since $x\cdot\!k=0$, the reduced energy is
\begin{equation}
\bar E=\frac12\Omega^t\tilde I\Omega+mgs\cdot\nclM.
\end{equation}

\note[Verify the reduced equations of motion]\notetarget{nt:verify-reduced-eq}
\iftoggle{vector-field-to-arxiv}
{To verify the reduced vector field, using}
{Using}
a multiplier~$\lambda$ for the constraint~\eqref{eq:reduced-K-constraint}
to $\bar K$ (and remembering 
that~$\deltaA_i=\nclM\times\LclM\,\deltas_i$), the equations for the
reduced vector field are
\begin{equation}\begin{split}\label{eq:reduced-euler-lagrange}
  &(\tilde I\,\deltaA_1)\cdot\delta\Omega_2
  -(\tilde I\,\deltaA_2)\cdot\delta\Omega_1
  +(\tilde I\Omega)\cdot(\deltaA_1\times\deltaA_2)
  +m(\Omega\times\deltas_1)\cdot(s\times\deltaA_2)
  -m(\Omega\times\deltas_2)\cdot(s\times\deltaA_1)
\\
  &\qquad
  =(\tilde I\Omega)\cdot\delta\Omega_2
  -m(s\times\Omega)\cdot(\Omega\times\deltas_2)
  -mg(s\times A^{-1}\!k)\cdot\deltaA_2
\\
  &\qquad\qquad\mbox{}
  +\lambda\bigl(
  \nclM^t\tilde I\,\delta\Omega_2
  -\Omega^t\tilde I\LclM\,\deltas_2
  +m(\nclM\cdot\Omega)s^t\,\deltas_2\bigr),
\\
  &\nclM^t\tilde I\,\delta\Omega_1
  -\Omega^t\tilde I\LclM\,\deltas_1
  +m(\nclM\cdot\Omega)s^t\,\deltas_1=0,
\end{split}\end{equation}
where $\deltas_2\in\TclM$ and $\delta\Omega_2\in\bbR^3$ are arbitrary. Setting
$\deltas_2=0$ (so then $\deltaA_2=0$), \eqref{eq:reduced-euler-lagrange} become
\begin{equation}\begin{split}\label{eq:vector-field-reduced-delta-s-zero}
  &(\tilde I\,\deltaA_1)\cdot\delta\Omega_2
  =(\tilde I\Omega)\cdot\delta\Omega_2
  +\lambda\,\nclM\cdot\tilde I\,\delta\Omega_2,
\\
  &\nclM^t\tilde I\,\delta\Omega_1
  -\Omega^t\tilde I\LclM\,\deltas_1
  +m(\nclM\cdot\Omega)s^t\,\deltas_1=0.
\end{split}\end{equation}
From the first~$\nclM\times \LclM\,\deltas_1-\Omega=\lambda\,\nclM$,
because $\tilde I\,\delta\Omega_2$ is arbitrary. Then the cross-product with
$\nclM$ provides $-\LclM\,\deltas_1-\nclM\times\Omega=0$, i.e.,
$\LclM\,\deltas_1=\Omega\times\nclM$, which is the $ds/dt$ equation
of \eqref{eq:vector-field-reduced}, while the dot-product with $\nclM$
obtains $\lambda=-\nclM\cdot\Omega$. Noting that
\begin{equation*}
  \nclM\cdot\bigl(s\times(\deltas_1\times\Omega)\bigr)
  =
  \nclM\cdot\bigl(
  (s\cdot\Omega)\deltas_1-(s\cdot\deltas_1)\Omega)\bigr)
  =
  -(\nclM\cdot\Omega)(s\cdot\deltas_1),
\end{equation*}
the second equation of~\eqref{eq:vector-field-reduced-delta-s-zero} is
\begin{equation*}
  \nclM\cdot\bigl(
  \tilde I\,\delta\Omega_1
  -(\tilde I\Omega)\times\Omega
  -ms\times(\deltas_1\times\Omega)\bigr)=0
\end{equation*}
corresponding to the $\nclM$ component of the~$d\Omega/dt$ equation
of~\eqref{eq:vector-field-reduced}. For the component orthogonal to $\nclM$,
assuming~$\deltas_2$ is arbitrary and setting
$\delta\Omega_2=0$, \eqref{eq:reduced-euler-lagrange}~becomes
\begin{equation}\begin{split}\label{eq:vector-field-reduced-delta-omega-zero}
  &-(\tilde I\,\deltaA_2)\cdot\delta\Omega_1 +(\tilde I\Omega)\cdot(\delta
  A_1\times\deltaA_2) +m(\Omega\times\deltas_1)\cdot(s\times\deltaA_2)
  -m(\Omega\times\deltas_2)\cdot(s\times\deltaA_1)
\\
  &\qquad
  =
  -m(s\times\Omega)\cdot(\Omega\times\deltas_2)
  +mg(s\times\nclM)\cdot\deltaA_2
  -(\nclM\cdot\Omega)\bigl(
    -\Omega^t\tilde I\LclM\,\deltas_2
  +m(\nclM\cdot\Omega)s^t\,\deltas_2\bigr).
\end{split}\end{equation}
But 
$\deltaA_1
  =\nclM\times\LclM\,\deltas_1
  =\nclM\times(\Omega\times\nclM)
  =\Omega-(\nclM\cdot\Omega)\nclM
$,
so the second term on the left
of~\eqref{eq:vector-field-reduced-delta-omega-zero} is
\begin{equation}\begin{split}
  (\tilde I\Omega)\cdot(\deltaA_1\times\deltaA_2)
  &=
  \bigl((\tilde I\Omega)\times(\Omega-(\nclM\cdot\Omega)\nclM)\bigr)
  \cdot\deltaA_2
\\
  &=
  (\tilde I\Omega)\times\Omega
  -(\nclM\cdot\Omega)\bigl((\tilde I\Omega)\times\nclM\bigr)\cdot\deltaA_2,
\end{split}\end{equation}
while the fourth term on the left
of \eqref{eq:vector-field-reduced-delta-omega-zero} is
\begin{equation}\begin{split}
  &-m(\Omega\times\deltas_2)\cdot(s\times\deltaA_1)
\\
  &\qquad\qquad=
  -m(\Omega\times\deltas_2)\cdot\bigl(s\times
  (\Omega-(\nclM\cdot\Omega)\nclM)\bigr)
\\
  &\qquad\qquad=
  -m\bigl(s\times\Omega)\cdot(\Omega\times\deltas_2)
  +m(\nclM\cdot\Omega)\bigl(
  (s\cdot\Omega)\cdot(\nclM\cdot\deltas_2)
  -(\nclM\cdot\Omega)(\deltas_2\cdot s)\bigr)
\\
 &\qquad\qquad=
  -m\bigl(s\times\Omega)\cdot(\Omega\times\deltas_2)
  -m(\nclM\cdot\Omega)^2(\deltas_2\cdot s).
\end{split}\end{equation}
So \eqref{eq:vector-field-reduced-delta-omega-zero} is
$
  \bigl(-(\tilde I\delta\Omega_1)
  +(\tilde I\Omega)\times\Omega
  +(m(\Omega\times\deltas_1)\times s)\bigr)\cdot\deltaA_2
  =
  mg(s\times\nclM)\cdot\deltaA_2
$,
i.e., the component of \eqref{eq:vector-field-reduced} orthogonal
to~$\nclM$.

{\footnotesize\ifdef\makebiblio%
{\bibliography{0one-rolling-body.bib}}%
{\expandafter
}}

\end{document}